\documentclass[aps,twocolumn]{aastex62}
\usepackage{amsmath, amssymb, graphicx, color}
\def\aap{Astron. Astrophys.}
\def\aj{Astron.J.}
\def\mnras{MNRAS}
\def\modified{} 

\begin{document}
\title{Supernova Magnitude Evolution and PAge Approximation}
\author{Zhiqi Huang}
\email{huangzhq25@mail.sysu.edu.cn}
\affiliation{School of Physics and Astronomy, Sun Yat-Sen University, 2 Daxue Road, Tangjia, Zhuhai, 519082, P.R.China}
\correspondingauthor{Zhiqi Huang}

\date{\today}
\begin{abstract}
  The evidence of environmental dependence of Type Ia supernova luminosity has inspired recent discussion about whether the late-universe cosmic acceleration is still supported by supernova data. We adopt the $\Delta\mathrm{HR}/\Delta\mathrm{age}$ parameter, which describes the dependence of supernova absolute magnitude on the age of supernova progenitor, as an additional nuisance parameter. Using the Pantheon supernova data, a lower bound $\ge 12\,\mathrm{Gyr}$ on the cosmic age, and a Gaussian prior $H_0 = 70\pm 2\,\mathrm{km\,s^{-1}Mpc^{-1}}$ on the Hubble constant, we reconstruct the cosmic expansion history. Within the flat $\Lambda$ cold dark matter ($\Lambda$CDM) framework, we still find a $5.6\sigma$ detection of cosmic acceleration. This is because a matter dominated decelerating universe would be too young to accommodate observed old stars with age $\gtrsim 12\,\mathrm{Gyr}$. A decelerating but non-flat universe is marginally consistent with the data, however, only in the presence of a negative spatial curvature $\sim$ two orders of magnitude beyond the current constraint from cosmic microwave background data. Finally, we propose a more general Parameterization based on the cosmic Age (PAge), which is {\it not} directly tied to the dark energy concept and hence is ideal for a null test of the cosmic acceleration. We find that, for a magnitude evolution rate $\Delta\mathrm{HR}/\Delta\mathrm{age} \lesssim 0.3\,\mathrm{mag}/5.3\,\mathrm{Gyr}$ \citep{Kang20}, a spatially flat and decelerating PAge universe is fully consistent with the supernova data and the cosmic age bound, and has no tension with the geometric constraint from the observed CMB acoustic angular scales. 
\end{abstract}



\section{Introduction}\label{sec:intro}

The accelerated expansion of the late universe, one of the greatest puzzles of modern physics, was firstly indicated by the ``unexpected extra dimming'' of high-redshift Type Ia supernovae~\citep{Riess98, Schmidt98, Perlmutter99, Pantheon}. It is explained in modern cosmology by a hypothetical dark energy component, whose microscopic nature is often interpreted as a cosmological constant $\Lambda$ ($\Lambda$CDM model where CDM stands for cold dark matter), or an unknown fluid component with a negative equation of state $w$ ($w$CDM model). Within the $\Lambda$CDM or $w$CDM framework, the late-time cosmic acceleration is also confirmed by a few independent cosmological probes, such as the cosmic microwave background (CMB)~\citep{Planck2018Params} and the baryon acoustic oscillations (BAO)~\citep{BAO-SDSS-DR12-LOWZ, BAO-SDSS-DR12-CMASS}. However, CMB and BAO constraints are more model dependent, as the prediction of observables depends not only on the expansion history of the universe, but also on the growth of inhomogeneities that depends on more detailed properties of cosmic ingredients.

If the luminosity of Type Ia supernovae can be calibrated to be a constant (with small and unbiased scattering), the Hubble diagram of supernovae would be the currently most direct and model-independent evidence for cosmic acceleration. The empirical standardization of supernova peak luminosity is obtained by a calibration against the light-curve shape and the color.  The universality of such a standardization procedure is based on the assumption that Type Ia supernova explosion is triggered by a critical condition that has little to do with its galactic environment and the past history of its progenitor. This assumption has been intensively investigated in the past decade. A series of work has found correlation between the standardized absolute magnitude of Type Ia supernovae and properties of their host galaxies~\citep{Hicken09, Sullivan10, Rigault13, Rigault15, Rigault18, Roman18, Kim19}. In particular, Ref.~\citep{Kim19} claimed a detection of a correlation between standardized supernova luminosity and stellar population age at a 99.5\% confidence. In a subsequent work, the authors interpret the correlation as a $\sim 0.27\,\mathrm{mag}/5.3\,\mathrm{Gyr}$  dependence of supernova standarized magnitude on the age of its progenitor~\citep{Kang20}. Because the typical age of supernova progenitor decreases with redshift, such a dependence dims supernovae at high redshift, and therefore is observationally degenerate with the believed accelerating expansion of the universe. For a $\sim 0.27\,\mathrm{mag}/5.3\,\mathrm{Gyr}$ supernova magnitude evolution, Ref.~\citep{Kang20} showed a concrete example that the supernova Hubble diagram can be roughly fit by an open CDM universe without late-time acceleration. However, the spatial curvature parameter $\Omega_k = 0.73$ used in the example is $\sim$ two orders of magnitude beyond the CMB constraint~\citep{Planck2018Params}. Although the CMB constraint on the flatness of universe is model-dependent, a $\sim$ two orders of magnitude boost, if ever possible, may require very careful (and fine-tuned) constrution of model.

It might be puzzling why a large $\Omega_k$ is needed to fit the Hubble diagram, if supernova magnitude evolution already (at least qualitatively) mimics the effect of $\Lambda$. The real problem for a decelerating universe without $\Omega_k$ is not the detailed quantitative difference in the Hubble diagrams, but the cosmic age! A successful cosmological model must predict a cosmic age  $t_0>t_{\star}$, where $t_{\star}$ is the maximum age for the oldest stars. The currently most accurate astrophysical determination of $t_{\star}$ is based on the separation of isochrones of different ages on the HR diagram around the turn-off and subgiant branch. The recently improved parallaxes and spectra by the Hubble space telescope and the Gaia spacecraft give an estimation $t_{\star} \gtrsim 12\,\mathrm{Gyr}$, with the uncertainty reduced to a sub-Gyr level~\citep{Vandenberg2014, GAIA_Age, AgeReview}. Thus, a flat CDM universe, which predicts a cosmic age $\sim 9\,\mathrm{Gyr}$, does not pass the astrophysical tests.

In summary, the supernova magnitude evolution, if confirmed by further observations, will have a significant impact on low-redshift cosmology, but it is yet unclear whether a non-accelerating universe can be made consistent with the observational facts. In this work, we will extend the qualitative discussion in Ref.~\citep{Kang20} to a full quantitative Bayesian exploration of  the cosmological implication of supernova magnitude evolution. We will start with the non-flat $\Lambda$CDM model, and proceed to a more general framework beyond the usual concept of dark energy.

Throughout the article we use natural units $c=\hbar=1$. A dot denotes the derivative with respect to the cosmological time $t$. The scale factor $a$ of the Friedmann-Robertson-Walker metric is normalized to unity today.

\section{Non-flat $\Lambda$CDM Revisited}

We use the Pantheon supernova catalog~\citep{Pantheon} and  modify its likelihood by adding a progenitor age modulated magnitude
\begin{equation}
  \Delta m = \frac{\Delta\mathrm{HR}}{\Delta\mathrm{age}} \times \tau_{\rm median}(z),
\end{equation}
where $\tau_{\rm median}(z)$ is the median value of the progenitor age $\tau$ for an observed supernova at redshift $z$.
We assumes the following priors: $0 \le \Delta\mathrm{HR}/\Delta\mathrm{age} < 0.3\,\mathrm{mag}/5.3\,\mathrm{Gyr}$,  $H_0 = 70\pm 2 \,\mathrm{km\,s^{-1}Mpc^{-1}}$ (Gaussian), and the age of universe $t_0 > 12\,\mathrm{Gyr}$. The median age of supernova progenitor is computed with the following probability density of finding a supernova at redshift $z$ with progenitor age $\tau$:
\begin{equation}
  P(\tau; z)d\tau \propto \left\{
  \begin{array}{ll}
    \frac{\tau^{\alpha} d\tau}{(t_p^\alpha + \tau^{\alpha-s}t_p^s)\left[10^{A(z'-z_0)}+10^{B(z'-z_0)}\right]}, & \text{if}\ t(z)>\tau \\
    0, & \text{otherwise}
  \end{array}
  \right.  \label{eq:prob}
\end{equation}
where $t_p=0.2\,\mathrm{Gyr}$, $z_0=1.243$, $A=-0.997$, $B=0.241$, and $z'$ is the redshift that satisfies $t(z)-t(z') = \tau$. More about the details of the recipe of supernova progenitor age can be found in \citet{Kang20} and references therein.

The purpose of the simple exercise done here is to compare with \citet{Kang20} from a theoretical perspective. Eq.~\eqref{eq:prob} should not be overly interpreted as a thorough and accurate study on the Pantheon samples. Calibration to the actual ages of the stellar systems that make up the Pantheon supernova sample, which ideally should be done, is non-trivial and beyond the scope of this tentative exploration. Noticeably, while this work is under review, a new analysis of the supernova samples used in~\citet{Kang20} claimed no evidence for supernova luminosity evolution after removal of a single {\modified poorly-sampled} supernova~\citep{Rose2020}. {\modified Neither did \citet{Rose2020} find any significant residual host-age dependence for Pantheon samples after standardization.}

\begin{figure}
  \includegraphics[width=\columnwidth]{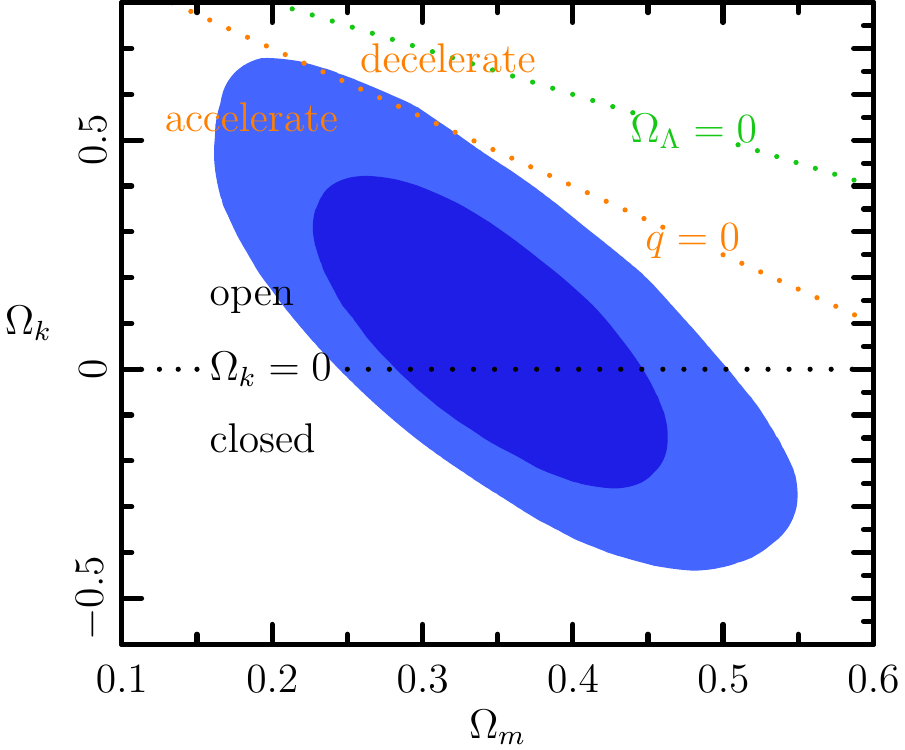}
  \caption{The marginalized 68.3\% (dark blue) and 95.4\% (light blue) confidence-level contours for the non-flat $\Lambda$CDM model. \label{fig:ommomk}}
\end{figure}

We first perform Monte Carlo Markov Chain (MCMC) analysis for the flat and non-flat $\Lambda$CDM models, respectively. The results are summarized in Fig.~\ref{fig:ommomk}. For flat $\Lambda$CDM model, the posterior of the deceleration parameter $q_0= -0.45\pm 0.08$ gives a $\sim 5.6\sigma$ detection of cosmic acceleration. For the non-flat $\Lambda$CDM model, a decelerating universe is marginally consistent with data, however, at the price of introducing an enormously large $\Omega_k\gtrsim 0.5$, which is strongly disfavored by the CMB data~\citep{Planck2018Params}. 

The example in \citet{Kang20}, $(\Omega_m, \Omega_\Lambda)=(0.27, 0)$, is well outside the $2\sigma$ contour. In fact, the entire $\Omega_\Lambda=0$ line is disfavored by the data, mainly because we have used a cosmic age prior $t_0> 12\mathrm{Gyr}$, which was not considered in \citet{Kang20}.

\section{The PAge Approximation \label{sec:wavelet}}

The main purpose of this work is to explore the possibility of a non-accelerating universe. A proper null test of cosmic acceleration should be done in a framework beyond the concept of dark energy. We propose a very simple, yet powerful parameterization based on the cosmic Age, which we dub as PAge approximation. The PAge approximation contains the same number of parameters as $w$CDM, but covers a broader class of scenarios beyond the usual dark energy concept.

Since the early 2000's, a Taylor expansion of the luminosity distance $d_L$ as a function of the redshift~\citep{Jerk}
\begin{equation}
  d_L(z) \approx \frac{z}{H_0}\left[1+\frac{1}{2}(1-q)z-\frac{1}{6}\left(1-q-3q^2+j-\Omega_k\right)z^2\right].~\label{eq:jerk}
\end{equation}
where the ``jerk'' parameter $j = \left.\frac{a^2\dddot a}{\dot a^3}\right\vert_{z=0}$, has been widely used in the literature for model-independent explorations beyond the dark energy concept. While the Taylor expansion of $z$ is convenient and is suitable for supernova data analysis at low redshift, it may fail at $z\gtrsim 1$ that is well accessible by a modern supernova catalog. Moreover, physical conditions such as $\frac{d d_L}{dz}>0$ (distance increases with redshift) and $\frac{dH}{dz}>0$ (background energy density decreases with time) are complicated in the $q$-$j$ space. Finally, because the Taylor approximation contains no information about high-redshift universe, the cosmic age or the distance to the last scattering surface of CMB are incomputable with Eq.~\eqref{eq:jerk}. The ``jerk'' parameterization by design is immune to any high-redshift criticism.

Because of the aforementioned disadvantages of the local Taylor expansion, we propose instead a global approximation of the cosmic expansion history. The PAge approximation is based on two assumptions: (i) the universe is dominated by matter at high redshift $z\gg 1$; (We ignore the radiation component and the very short period before matter domination.) (ii) the product of the cosmological time $t$ and the Hubble expansion rate $H$ can be approximated as a quadratic function of $t$. It can be easily shown that the two assumptions lead to
\begin{equation}
  \frac{H}{H_0} = 1 + \frac{2}{3}\left(1-\eta\frac{H_0t}{p_{\rm age}}\right)\left(\frac{1}{H_0t} - \frac{1}{p_{\rm age}}\right),
\end{equation}
where $p_{\rm age} = H_0t_0$ is the product of Hubble constant $H_0$ and the current age of the universe $t_0$, and the phenomenological parameter $\eta $ can be regarded as a quadratic fitting parameter.

There are immediately some advantages of using the PAge approximation. For instance, both of the physical conditions $\frac{dd_L}{dz}>0$ and $\frac{dH}{dz}>0$ can be guaranteed by a simple bound $\eta<1$, and global quantities such as the observational bounds on the cosmic age can be easily applied to PAge. More importantly, the popular models in the literature - flat or non-flat, $\Lambda$CDM or $w$CDM models - can all be approximately mapped to the PAge space by matching the age of universe $t_0$ and the current deceleration parameter $q_0 = -\frac{a\ddot a}{\dot a^2}$. A few examples are given in Table~\ref{tab:pageapp}. 
\begin{table}
  \caption{PAge approximations \label{tab:pageapp}}
  \begin{tabular}{l l l l l l}
    \hline
    \hline
    $\Omega_m$ & $w$ & $\Omega_k$ &  $p_{\rm age}$ & $\eta$ & max.$|\Delta \mu|$ \\
    \hline
    $0.3$ & $-1$ & $0$ & $0.964$ & $0.373$ & $9.9\times 10^{-3}$ \\
    $0.5$ & $-1$ & $0$ & $0.831$ & $0.223$ & $3.3\times 10^{-3}$ \\
    $1$ & $-1$ & $0$ & $2/3$ & $0$ & $0$ \\
    $0.3$ & $-1$ & $0.1$ & $0.935$ & $0.279$ & $4.2\times 10^{-3}$ \\
    $0.3$ & $-1$ & $-0.1$ & $0.997$ & $0.479$ & $1.6\times 10^{-3}$ \\
    $0.3$ & $-0.9$ & $0$ & $0.948$ & $0.251$ & $1.5\times 10^{-2}$ \\
    $0.3$ & $-1.1$ & $0$ & $0.978$ & $0.505$ & $1.3\times 10^{-2}$ \\    
    \hline
  \end{tabular}
\end{table}
As shown in the last column of Table~\ref{tab:pageapp}, such a mapping typically yields $\lesssim 1\%$ errors in distance modulus $\mu$ at $z<1.5$, which are negligible for current supernova data analyses. For future high-precision supernova cosmology, however, the difference between physically motivated models and PAge approximation may require more careful treatment. 

\begin{figure}
  \includegraphics[width=\columnwidth]{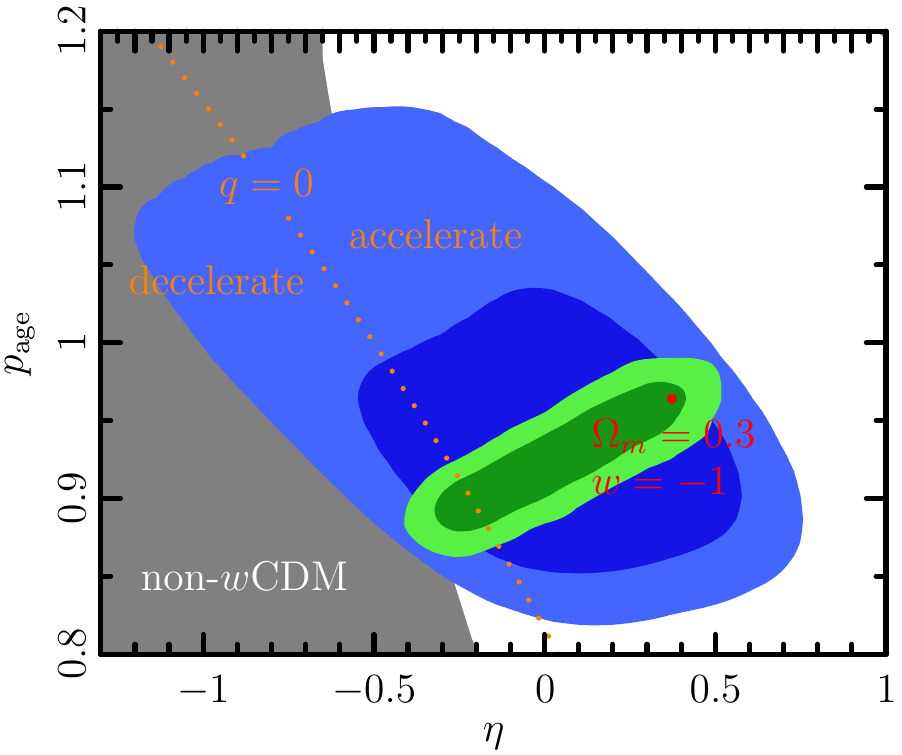}
  \caption{The marginalized 68.3\% and 95.4\% confidence-level contours for the flat PAge model. The dark blue and light blue contours are obtained with Pantheon supernova data, cosmic age prior and $H_0$ prior, while an additional CMB distance prior is added to produce the dark green and light green contours. Each point in the white background region can be approximately mapped to a $w$CDM model. The gray background color shows the parameter space beyond $w$CDM. The red dot corresponds to $(\eta, p_{\rm age})=(0.373, 0.964)$, which is a good approximation of the concordance $\Lambda$CDM ($\Omega_m=0.3$).\label{fig:page}}
\end{figure}

The loss of $\sim 1\%$ accuracy in distance modulus is compensated by an unexplored beyond-$w$CDM parameter space, as shown by the gray background color in Fig.~\ref{fig:page}. Since the mapping between PAge and an effective dark energy model is approximate and the approximation becomes worse towards the gray region, the division between $w$CDM and non-$w$CDM is only in a approximate sense. Roughly speaking, the white region can be approximated with non-interacting dark energy models, while the gray region represents more complicated models, such as an interacting dark component that exchanges energy with CDM. In general, each point on the $\eta$-$p_{\rm age}$ plane should be regarded as an approximation of many physical models that share a similar expansion history.

The marginalized constraints on $\eta$ and $p_{\rm age}$ in Fig.~\ref{fig:page} are obtained for a flat PAge universe with supernova magnitude evolution $0 \le \Delta\mathrm{HR}/\Delta\mathrm{age} < 0.3\,\mathrm{mag}/5.3\,\mathrm{Gyr}$, the Hubble constant $H_0 = 70\pm 2 \,\mathrm{km\,s^{-1}Mpc^{-1}}$, and the cosmic age $t_0 > 12\,\mathrm{Gyr}$. For the dark green and light green contours we have used an additional prior on the comoving distance to the last scattering surface ($13.8\mathrm{Gpc} < \left.d_A^{\rm com}\right\vert_{z=1089} < 14\mathrm{Gpc}$) to guarantee that the theory is roughly consistent with observed CMB acoustic angular scales. The results shown in Fig.~\ref{fig:page} suggest that a decelerating PAge universe can fit the supernova data very well without obvious tension with CMB observations.

\section{Discussion and Conclusions}

The observational hints of supernova magnitude evolution may challenge the late-time acceleration and the standard flat $\Lambda$CDM paradigm. \citet{Kang20} proposed that a non-flat universe without dark energy may roughly fit the supernova data. We did a full Bayesian analysis in this work and showed that when a cosmic age bound is applied, (i) a non-flat $\Omega_\Lambda=0$ universe is inconsistent with the data, mainly due to the cosmic age bound; (ii) a decelerating non-flat $\Lambda$CDM universe is marginally consistent with the data, but it requires an enormously large $\Omega_k\gtrsim 0.5$ that can hardly be made consistent with CMB observations.

The $w$CDM model is another popular extension of $\Lambda$CDM. The philosophy of $w$CDM or its extensions with time-dependent $w$ is to assume simplicity in the dark energy equation of state, which may be reasonable if the dark energy concept is accepted a priori. For a null test of the cosmic acceleration, however, we need a more general description beyond the dark energy concept.

The PAge approximation is a different philosophy. It assumes simplicity in $Ht$ rather than in dark energy $w$. PAge is a phenomenological parameterization without specifying the underlying physics that drives the late-time expansion of universe. Thus, a full calculation of CMB and BAO observables requires further model constructions. Nevertheless, we assume that the flatness and acoustic angular scale constraint from CMB will remain roughly valid. In this context we find that a decelerating PAge universe is fully consistent with the data, whereas the concordance $\Lambda$CDM shown as a red dot in Fig.~\ref{fig:page} is nothing but a good fit on the edge of $1\sigma$ contour. The coincidental proximity $p_{\rm age}\approx 1$ in the $\Lambda$CDM framework has inspired some recent discussion about whether we are living in a special cosmic era~\citep{Avelino2016}. In the much more flexible PAge framework, the viable range of $p_{\rm age}$ is relaxed to $\sim [0.86, 1.00]$ (99.7\% confidence), with the lower bound mainly from the astrophysical constraint and the upper bound mainly from CMB. It would be interesting to see whether the future improved astrophysical observations will push $p_{\rm age}$ toward $\approx 1$ in the PAge framework.

The BAO standard ruler inferred from the wiggling of galaxy power spectrum in principle can be used to constrain the background expansion of the universe, too. However, redshift-space distortion (RSD) and nonlinear structures in the late universe can bias the location of BAO peaks in a model dependent way. RSD and nonlinear corrections for models far beyond the concordance $\Lambda$CDM can be very non-trivial. We leave the PAge exploration of BAO, as well as of many other potential probes~\citep{Wei2016,Zheng2019,Zhang2019,H0LiCow,STRIDES2019}, as our future work.


\end{document}